\pdfoutput=1
\documentclass[prd,showpacs,letterpaper,10pt,aps,notitlepage]{revtex4}

\usepackage{amsfonts,amsmath,graphicx,bm}

\newcommand{\nb}{\phantom{0}}
\newcommand{\bs}[1]{\ensuremath{{\boldsymbol{#1}}}}

\def\sfrac#1#2{{\textstyle\frac{#1}{#2}}}

\begin{document}

\title{Prediction of the $\bs{\Omega_{bbb}}$ mass from lattice QCD}

\author{Stefan Meinel}
\affiliation{Department of Physics, College of William \&  Mary, Williamsburg,
  VA 23187-8795, USA}

\date{October 27, 2010}


\begin{abstract}
The mass of the triply heavy baryon $\Omega_{bbb}$ is calculated in lattice QCD with $2+1$ flavors of light sea quarks.
The $b$ quark is implemented with improved lattice NRQCD. Gauge field ensembles from both the RBC/UKQCD and MILC
collaborations with lattice spacings in the range from 0.08 fm to 0.12 fm are used. The final result for the $\Omega_{bbb}$
mass, which includes an electrostatic correction, is $ 14.371 \pm 0.004_{\rm \:stat} \pm 0.011_{\rm \:syst} \pm 0.001_{\rm \:exp}$ GeV.
The hyperfine splitting between the physical $J=3/2$ state and a fictitious $J=1/2$ state is also calculated.
\end{abstract}

\maketitle

\section{Introduction}

Hadrons containing only heavy valence quarks play a key role in the study of QCD due to their approximately nonrelativistic
nature and their clean spectrum of narrow states. Many precise experimental results are available for heavy quarkonium mesons,
and their analysis has contributed significantly to the understanding of the quark-antiquark forces \cite{Bali:2000gf, Brambilla:2004wf}.
However, the nature of the QCD gauge symmetry is exhibited in the most direct way in the \emph{baryons},
where the number of valence quarks is equal to the number of colors. It is therefore of great interest to explore triply-heavy baryons
like the $\Omega_{bbb}$. Masses and other properties of triply-heavy baryons have been calculated using various quark models
\cite{Ponce:1978gk, Hasenfratz:1980ka, Bjorken:1985ei, Tsuge:1985ei, Basdevant:1985ux, SchaffnerBielich:1998ci, Vijande:2004at,
Martin:1995vk, SilvestreBrac:1996bg, Migura:2006ep, Jia:2006gw, Faessler:2006ft, Martynenko:2007je, Gerasyuta:2007un, Roberts:2007ni, Bernotas:2008bu}
and sum rules \cite{Zhang:2009re}. Most potential-model calculations include only two-body forces, but QCD also leads to genuine three-body forces.
A three-body linear confinement term was found in the static three-quark potential calculated nonperturbatively from lattice QCD
\cite{Takahashi:2000te}. In perturbative QCD, three-body forces first arise at next-to-next-to-leading order \cite{Brambilla:2009cd}.

Triply-heavy baryons have not yet been observed experimentally (see \cite{Bjorken:1985ei, Saleev:1999ti,Baranov:2004er,GomshiNobary:2004mq, GomshiNobary:2005ur}
for experimental aspects and calculations of production rates in colliders), and the predictions of their properties can not yet be compared to the real world.
The model-dependent calculations can however be tested by comparing them to nonperturbative first-principles calculations in lattice QCD.

While there are several recent lattice QCD calculations of singly- and doubly-heavy baryon masses
\cite{Lewis:2008fu,Burch:2008qx,Na:2008hz,Detmold:2008ww,Lin:2009rx, Liu:2009jc},
there appears to be only one lattice result for a triply-heavy baryon mass in the literature so far:
the $\Omega_{ccc}$ in Ref.~\cite{Chiu:2005zc}. The calculation in \cite{Chiu:2005zc} was performed in the quenched approximation,
i.e.~neglecting the effects of dynamical light quarks, and the result for $M_{\Omega_{ccc}}$ obtained there
may also have significant discretization errors \cite{Dong:2007da} due to the use of a relativistic action for the charm quark at the rather large value of
$a m_c=0.8$ (where $m_c$ is the bare charm quark mass and $a$ is the lattice spacing).

In the following, a precise calculation of the mass of the triply-bottom baryon $\Omega_{bbb}$ in lattice QCD with $2+1$ flavors of dynamical
light quarks is reported (a preliminary result was given in \cite{Meinel:2009vv}).
The hyperfine splitting between the $\Omega_{bbb}$ baryon with $J=3/2$ and a fictitious $J=1/2$ state composed of three heavy quarks with the
same mass as the $b$ quark is also calculated. The computations are performed at lattice spacings in the range from approximately 0.08 fm to 0.12 fm. At these
values of the lattice spacing, the $b$ quark can be implemented very accurately with lattice nonrelativistic QCD (NRQCD)
\cite{Thacker:1990bm, Lepage:1992tx, Meinel:2010pv}. Most of the calculations in this work are done on gauge field ensembles
that were generated by the RBC and UKQCD collaborations using the Iwasaki action for the gluons and a domain-wall action for the sea quarks
\cite{Allton:2007hx,Allton:2008pn}. As a test of universality, calculations are also performed on gauge field ensembles generated
by the MILC collaboration using the L\"{u}scher-Weisz action for the gluons and a rooted staggered action for the sea quarks \cite{Bazavov:2009bb}.
The details of the gauge field ensembles are given in Sec.~\ref{sec:parameters}. The mass of the $\Omega_{bbb}$ is obtained by calculating the difference
$E_{\Omega_{bbb}}-\frac38\left( E_{\eta_b}+3 E_{\Upsilon}\right)$, as described in Sec.~\ref{sec:methods}. This quantity has only a weak dependence
on the $b$ quark mass and on the light quark masses, as demonstrated by the lattice results in Sec.~\ref{sec:results}. In the final result for
$M_{\Omega_{bbb}}$, a correction due to the electrostatic Coulomb interaction of the $b$ quarks is included (Sec.~\ref{sec:mass}).

\section{\label{sec:parameters}Lattice actions and parameters}

The problem with heavy quarks on the lattice is that relativistic actions develop large discretization
errors when the heavy-quark mass becomes comparable to the inverse lattice spacing. In this work, the $b$ quarks are therefore
implemented with lattice NRQCD \cite{Thacker:1990bm, Lepage:1992tx}, a nonrelativistic effective field theory
for heavy quarks. Unlike HQET \cite{Eichten:1989kb}, which is limited to heavy-light hadrons,
NRQCD can be applied to hadrons containing any number of heavy quarks. The use of NRQCD to study triply-heavy baryons
has already been suggested in \cite{Thacker:1990bm}.
The power counting in this case should be very similar to that for heavy quarkonia, and is given in terms
of the typical heavy-quark velocity inside the hadron. For bottomonium, one has $v^2\approx0.1$ (in units with $c=1$),
and the value is expected to be comparable for the $\Omega_{bbb}$ (the length scales are similar; see Eqs.~(\ref{eq:size_omega}) and (\ref{eq:size_upsilon}) in 
Sec.~\ref{sec:mass}).

The NRQCD action employed here includes all terms up to order $v^4$, as well as
Symanzik-improvement corrections (which reduce discretization errors). The matching to QCD
is performed at tree-level, and the action is tadpole-improved using the mean link $u_{0L}$
in Landau gauge \cite{Lepage:1992xa, Shakespeare:1998dt}.
The action is equal to the one used in \cite{Meinel:2010pv}, with $c_1=c_2=c_3=c_4=c_5=c_6=1$ and $c_7=c_8=c_9=0$ (in the notation used there.)
For comparison, some results obtained with the order-$v^2$ action, which has $c_5=c_6=1$ and
$c_1=c_2=c_3=c_4=c_7=c_8=c_9=0$, will also be given.

The details of the gauge field ensembles used in this work are shown in Table \ref{tab:lattices}. All ensembles
include the effects of dynamical up- down- and strange sea quarks (with $m_u=m_d$, in the following denoted as $m_l$).
The ensembles generated by the RBC and UKQCD collaborations \cite{Allton:2007hx,Allton:2008pn}
make use of a domain wall action \cite{Kaplan:1992bt,Shamir:1993zy,Furman:1994ky} for the sea quarks,
and the Iwasaki action \cite{Iwasaki:1983ck,Iwasaki:1984cj} for the gluons.
The domain wall action yields an exact chiral symmetry when the extent $L_5$ of the auxiliary fifth dimension is taken to infinity.
The Iwasaki action suppresses the residual chiral symmetry breaking at finite $L_5$ \cite{Aoki:2002vt}
(here, $L_5=16$).

Calculations are also performed on two ensembles generated by the MILC
collaboration \cite{Bazavov:2009bb}, using the AsqTad action \cite{Orginos:1998ue,Lepage:1998vj,Orginos:1999cr}
in combination with the rooting procedure \cite{Bazavov:2009bb} for the sea quarks,
and the tadpole-improved L\"uscher-Weisz action \cite{Luscher:1984xn, Luscher:1985zq, Alford:1995hw} for the gluons.
By comparing the results from the RBC/UKQCD and from the MILC ensembles, sea-quark and gluon discretization
effects can be disentangled from NRQCD discretization effects \cite{Meinel:2010pv}.

Table \ref{tab:lattices} also shows the lattice spacings of the ensembles, determined from the $\Upsilon(2S)-\Upsilon(1S)$ splitting
\cite{Meinel:2010pv}, and the values of the bare $b$-quark mass in lattice units, $a m_b$. The $b$-quark mass is set such that the
kinetic mass of the $\eta_b(1S)$ meson agrees with experiment within the statistical errors.
The main calculations in this work are performed directly at the physical values of
$a m_b$ as determined in \cite{Meinel:2010pv}; additional results for other values of $a m_b$ will also be given to illustrate
the dependence on $a m_b$. On the coarse MILC ensemble, the value of $a m_b$ used here is 0.9\% below the physical value obtained in
\cite{Meinel:2010pv}.

\begin{table*}[h!]
\begin{ruledtabular}
\begin{tabular}{lcclllcccl}
Collaboration & $L^3\times T$ & $\beta$ & $a m_l$ & $a m_s$  & $a m_b$ & $u_{0L}$ & $n_{\rm conf}$ & $a^{-1}$ (GeV) & $m_\pi$ (GeV) \\
\hline
RBC/UKQCD & $16^3\times 32$  & $2.13$ & $0.01$ & $0.04$     & $2.469$ & $0.8439$ & $352$ & $1.766(52)$ & $0.436(14)$  \\
          & $16^3\times 32$  & $2.13$ & $0.02$ & $0.04$     & $2.604$ & $0.8433$ & $355$ & $1.687(46)$ & $0.548(16)$  \\
          & $16^3\times 32$  & $2.13$ & $0.03$ & $0.04$     & $2.689$ & $0.8428$ & $711$ & $1.651(33)$ & $0.639(14)$  \\
\\[-1ex]
RBC/UKQCD & $24^3\times 64$  & $2.13$ & $0.005$ & $0.04$    & $2.487$ & $0.8439$ & $311$ & $1.763(27)$ & $0.3377(54)$  \\
          & $24^3\times 64$  & $2.13$ & $0.01$  & $0.04$    & $2.522$ & $0.8439$ & $266$ & $1.732(28)$ & $0.4194(70)$  \\
          & $24^3\times 64$  & $2.13$ & $0.02$  & $0.04$    & $2.622$ & $0.8433$ & $73$  & $1.676(42)$ & $0.541(14)$  \\
          & $24^3\times 64$  & $2.13$ & $0.03$  & $0.04$    & $2.691$ & $0.8428$ & $72$  & $1.650(39)$ & $0.641(15)$  \\
\\[-1ex]
RBC/UKQCD & $32^3\times 64$  & $2.25$ & $0.004$ & $0.03$    & $1.831$ & $0.8609$ & $314$ & $2.325(32)$ & $0.2950(40)$  \\
          & $32^3\times 64$  & $2.25$ & $0.006$ & $0.03$    & $1.829$ & $0.8608$ & $296$ & $2.328(45)$ & $0.3529(69)$  \\
          & $32^3\times 64$  & $2.25$ & $0.008$ & $0.03$    & $1.864$ & $0.8608$ & $270$ & $2.285(32)$ & $0.3950(55)$  \\
\\[-1ex]
MILC      & $24^3\times 64$  & $6.76$ & $0.005$ & $0.05$    & $2.64$  & $0.8362$ & $525$ & $1.647(14)$ & $0.460$ \\
\\[-1ex]
MILC      & $28^3\times 96$  & $7.09$ & $0.0062$ & $0.031$  & $1.86$  & $0.8541$ & $478$ & $2.291(22)$ & $0.416$ \\
\end{tabular}
\end{ruledtabular}
\caption{\label{tab:lattices}Summary of lattice parameters. The bare gauge couplings are given as $\beta=6/g^2$
(for the RBC/UKQCD ensembles) and $\beta=10/g^2$ (for the MILC ensembles). For the RBC/UKQCD ensembles, the
pion masses in lattice units were taken from \protect\cite{Allton:2007hx, Allton:2008pn, Syritsyn:2009np} and converted to physical units
using the lattice spacings given here. For the MILC ensembles, there are taste splittings between the different pions \protect\cite{Bazavov:2009bb},
and the root-mean-square masses taken from \protect\cite{Bazavov:2009tw,Bazavov:2009ir} are given.}
\end{table*}

\section{\label{sec:methods}Construction of two-point functions and analysis}

The ground-state energies of the $\eta_b$, the $\Upsilon$, and the $\Omega_{bbb}$ are extracted from fits to the Euclidean time-dependence
of suitable two-point functions at zero momentum. On a given lattice gauge field configuration,
the two-point functions for the $\eta_b$, the $\Upsilon$, and the $\Omega_{bbb}$ are defined as
\begin{eqnarray}
 C^{(\eta)}(t, t',\bs{x}') &=& \sum_{\bs{x}}
 G^{\dag ab}_{\alpha\beta}(t,\bs{x},t',\bs{x}') \:G_{\beta\alpha}^{ba}(t,\bs{x},t',\bs{x}'),  \\
 C^{(\Upsilon)}(t, t',\bs{x}') &=& \sum_{\bs{x}}
 (\gamma_j \gamma_5)_{\alpha\beta}\: G^{\dag ab}_{\beta\gamma}(t,\bs{x},t',\bs{x}') \:(\gamma_5\gamma_j)_{\gamma\delta}\: G_{\delta\alpha}^{ba}(t,\bs{x},t',\bs{x}'),  \\
 C^{(\Omega)}_{jk\:\alpha\delta}(t, t',\bs{x}') &=& \sum_{\bs{x}}
\epsilon_{abc}\:\epsilon_{fgh}\:(C\gamma_j)_{\beta\gamma}\:(\overline{C\gamma_k})_{\rho\sigma}
\: G_{\beta\sigma}^{af}(t,\bs{x},t',\bs{x}')\:G_{\gamma\rho}^{bg}(t,\bs{x},t',\bs{x}')\:G_{\alpha\delta}^{ch}(t,\bs{x},t',\bs{x}'). \label{eq:correlator_def}
\end{eqnarray}
Here, $a,b,c, ...$ are color indices (running from 1 to 3), $\alpha,\beta,\gamma,...$ are spinor indices (running from 1 to 4),
$C=\gamma_0\gamma_2$ is the (Euclidean) charge conjugation matrix, and the overline denotes
the Dirac conjugate. The indices $j,k$ are in the range from 1 to 3.
In the nonrelativistic gamma-matrix basis used here, the NRQCD heavy-quark propagator $G$ has vanishing lower spinor components (for $t>t'$):
\begin{equation}
 G(t,\bs{x},t',\bs{x}')=\left(\begin{array}{cc} G_\psi(t,\bs{x},t',\bs{x}') & 0 \\ 0 & 0 \end{array} \right), \label{eq:NRQCD_prop}
\end{equation}
where $G_\psi$ is the NRQCD propagator with two spinor components. The propagator $G_\psi$ may include at the source and/or sink
a Gaussian smearing operator
\begin{equation}
 \left(1 + \frac{r_S}{n_S}\Delta^{(2)}\right)^{n_S}, \label{eq:smear_op}
\end{equation}
where $\Delta^{(2)}$ is a covariant lattice Laplacian. The smearing is intended to improve the overlap with the ground state and reduce the contamination from excited states.
For the meson two-point functions, smearing is only performed on $G_\psi$, not $G_\psi^\dag$, while for the baryon two-point functions all three
$G_\psi$'s are treated equally.

\begin{figure}[h!]
 \centerline{\includegraphics[width=0.43\linewidth]{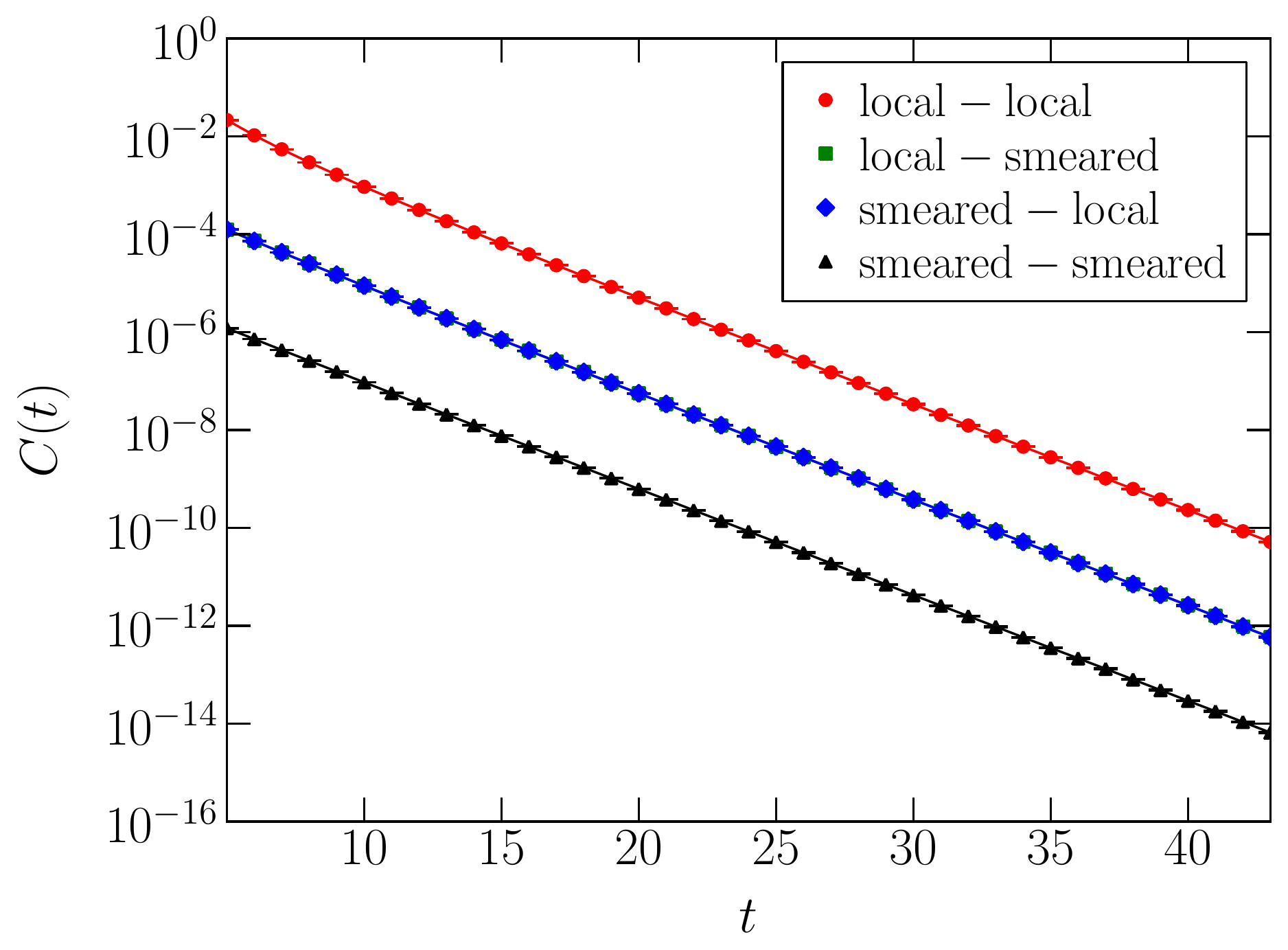}\hfill\includegraphics[width=0.43\linewidth]{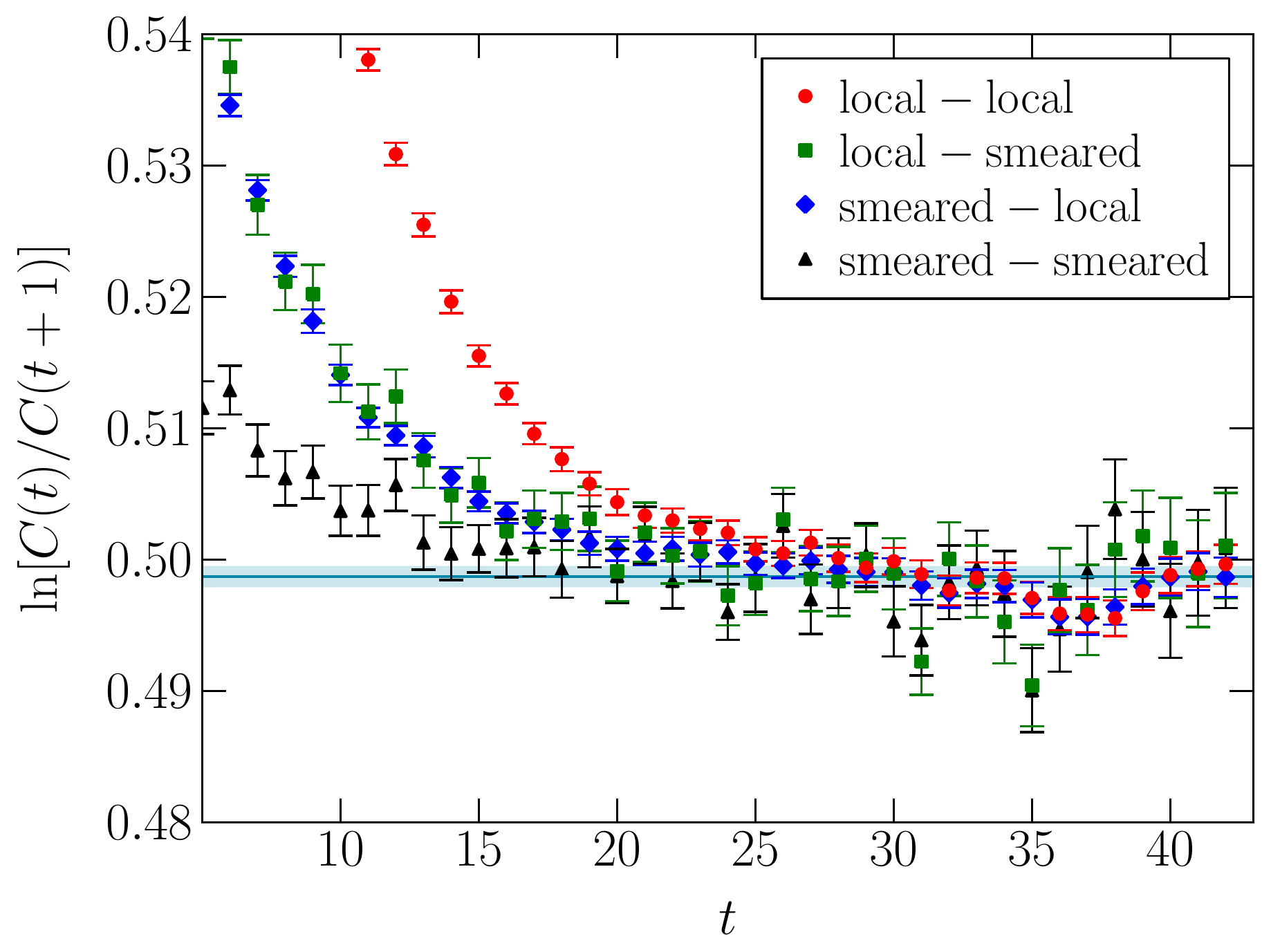}}
\caption{\label{fig:Omega_bbb_correlator}Left panel: matrix of $\Omega_{bbb}$ two-point functions $C(t)=\langle P^{(3/2)} \: C^{(\Omega)} \rangle$
with local and smeared interpolating fields (the local-smeared and smeared-local data coincide);
the lines are from a fit with 7 exponentials and $t_{\rm min}=5$.
Right panel: corresponding effective-energy plot and ground state energy.
The data are from the RBC/UKQCD ensemble with $L=32$, $a m_l=0.004$. Lattice units are used.}
\end{figure}

When defined through Eq.~(\ref{eq:correlator_def}), the two-point function $ C^{(\Omega)}_{jk\:\alpha\delta}(t, t',\bs{x}')$ couples to both the physical
spin-$3/2$ state $\Omega_{bbb}$, and an unphysical spin-$1/2$ state (when all three quark flavors are equal, this state violates the Pauli exclusion principle).
At large Euclidean time separation $t-t'$, the two-point function
(after averaging over gauge configurations) approaches the form
\begin{equation}
 C^{(\Omega)}_{jk} \rightarrow  Z_{3/2}^2\: e^{-E_{3/2}\:(t-t')}\:\sfrac12(1+\gamma_0)(\delta_{jk}-\sfrac13\gamma_j\gamma_k)
+ Z_{1/2}^2\: e^{-E_{1/2}\:(t-t')}\:\sfrac12(1+\gamma_0)\sfrac13\gamma_j\gamma_k,
\end{equation}
where $E_{3/2}=E_{\Omega_{bbb}}$ and $E_{1/2}$ are the ground-state energies of the $J=\frac32$ and $J=\frac12$ states, respectively (see for example \cite{Bowler:1996ws}).
The $J=\frac32$ and $J=\frac12$ contributions can be disentangled by multiplying with the projectors
\begin{eqnarray}
 P^{(3/2)}_{ij}&=&(\delta_{ij}-\sfrac13\gamma_i\gamma_j),\\
 P^{(1/2)}_{ij}&=&\sfrac13\gamma_i\gamma_j,
\end{eqnarray}
which gives
\begin{equation}
 P^{(J)}_{ij} \: C^{(\Omega)}_{jk} \rightarrow Z_J^2\: e^{-E_J\:(t-t')}\:\sfrac12(1+\gamma_0) P^{(J)}_{ik}.
\label{eq:baryon_spin_proj}
\end{equation}
Before the fitting, an average over all non-vanishing spinor- and Lorentz components is calculated, which is defined as
\begin{equation}
\langle P^{(J)} \: C^{(\Omega)} \rangle \:\:= \sum_{\begin{array}{c} i,k,\alpha,\delta, \\ {\scriptstyle \left[(1+\gamma_0)P^{(J)}_{ik}\right]_{\alpha\delta} \neq 0} \end{array}}
\frac{\left[P^{(J)}_{ij} \: C^{(\Omega)}_{jk}\right]_{\alpha\delta}}{\left[(1+\gamma_0)P^{(J)}_{ik}\right]_{\alpha\delta}}.
\end{equation}
In order to increase statistics, the two-point functions are
calculated for 32 different source locations $(t',\bs{x}')$ spread evenly across the lattice on each gauge field configuration.
The source locations are shifted randomly from configuration to configuration.
No significant autocorrelations were seen either between source locations or in molecular dynamics time.
An example of a matrix of $\Omega_{bbb}$ two-point functions is shown in Fig.~\ref{fig:Omega_bbb_correlator}.
The four different functions correspond to smeared or local sources and/or sinks. The data are fitted
by a sum of exponentials using the Bayesian technique from \cite{Lepage:2001ym}. The fit functions and
and priors are chosen as discussed in \cite{Meinel:2009rd}.

Due to the use of NRQCD, the energies extracted from fits of two-point functions contain a shift that is proportional
to the number of heavy quarks in the hadron. This shift cancels in the energy differences
\begin{equation}
 aE_{\Omega_{bbb}}-\frac38\left( aE_{\eta_b}+3 aE_{\Upsilon}\right) \label{eq:mdiff_spinav}
\end{equation}
and
\begin{equation}
aE_{\Omega_{bbb}}- \frac32aE_{\Upsilon}. \label{eq:mdiff_1s}
\end{equation}
As will be shown in Sec.~\ref{sec:results}, the quantity (\ref{eq:mdiff_spinav}), which contains the spin average of the $1S$ bottomonium
states, has a weaker dependence on the $b$-quark mass than (\ref{eq:mdiff_1s}) and is therefore preferred for the determination
of $M_{\Omega_{bbb}}$. The energy differences (\ref{eq:mdiff_spinav}) and (\ref{eq:mdiff_1s}) are computed using statistical
bootstrap in order to take into account correlations.

After conversion to physical units using the lattice spacings from the $\Upsilon(2S)-\Upsilon(1S)$ splittings,
the data from the RBC/UKQCD ensembles are extrapolated linearly in $m_\pi^2$ to the physical pion mass.
As in \cite{Meinel:2010pv}, the data from the $L=24$ and $L=32$ ensembles, which have the same
physical box size, are extrapolated simultaneously, allowing for an arbitrary dependence on the lattice spacing $a$.
Higher-order terms depending on both $a$ and $m_\pi$ are neglected.

\section{\label{sec:results}Lattice results}

\begin{figure*}[h!]
 \includegraphics[width=0.43\linewidth]{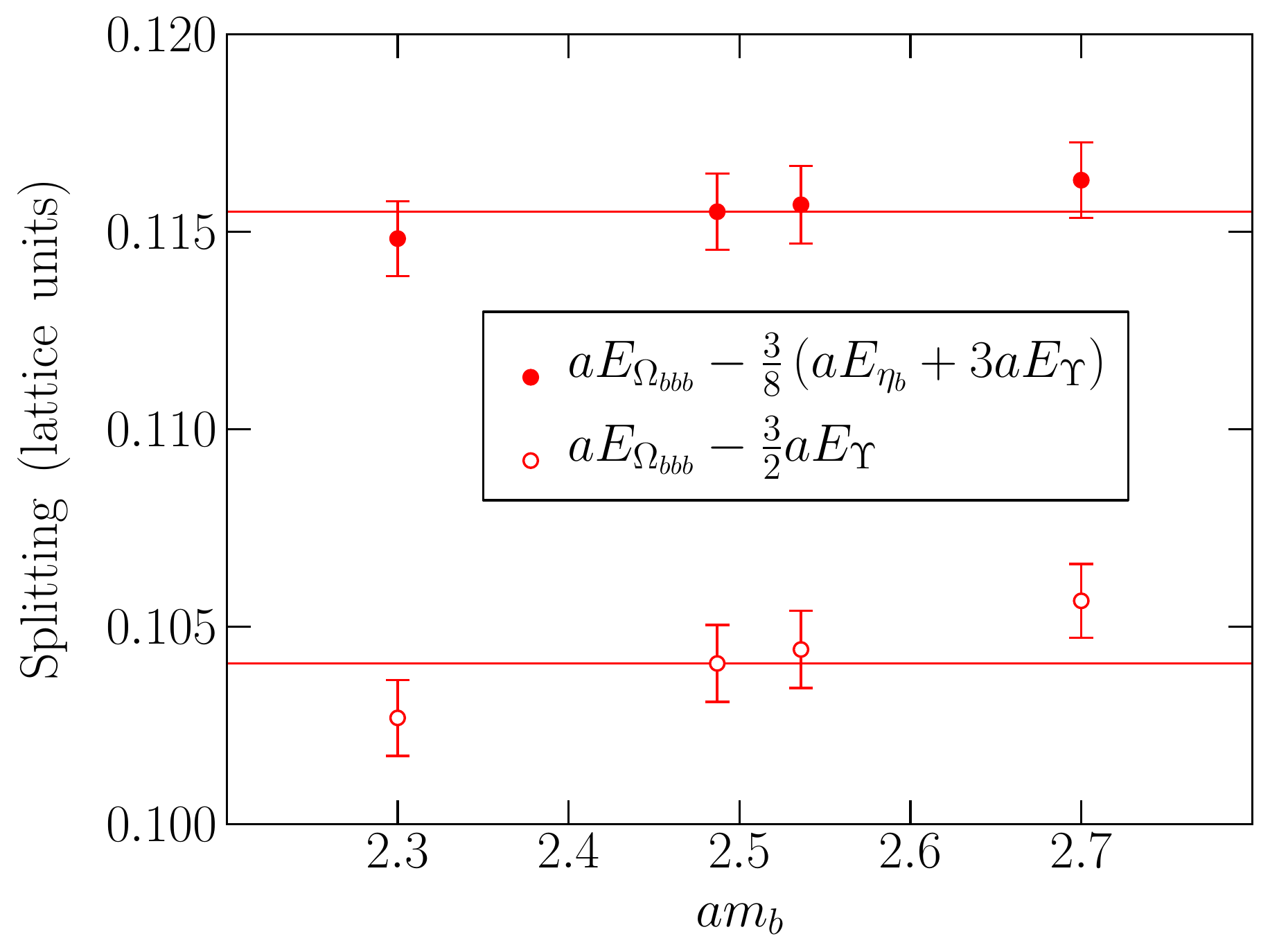} \hfill \includegraphics[width=0.43\linewidth]{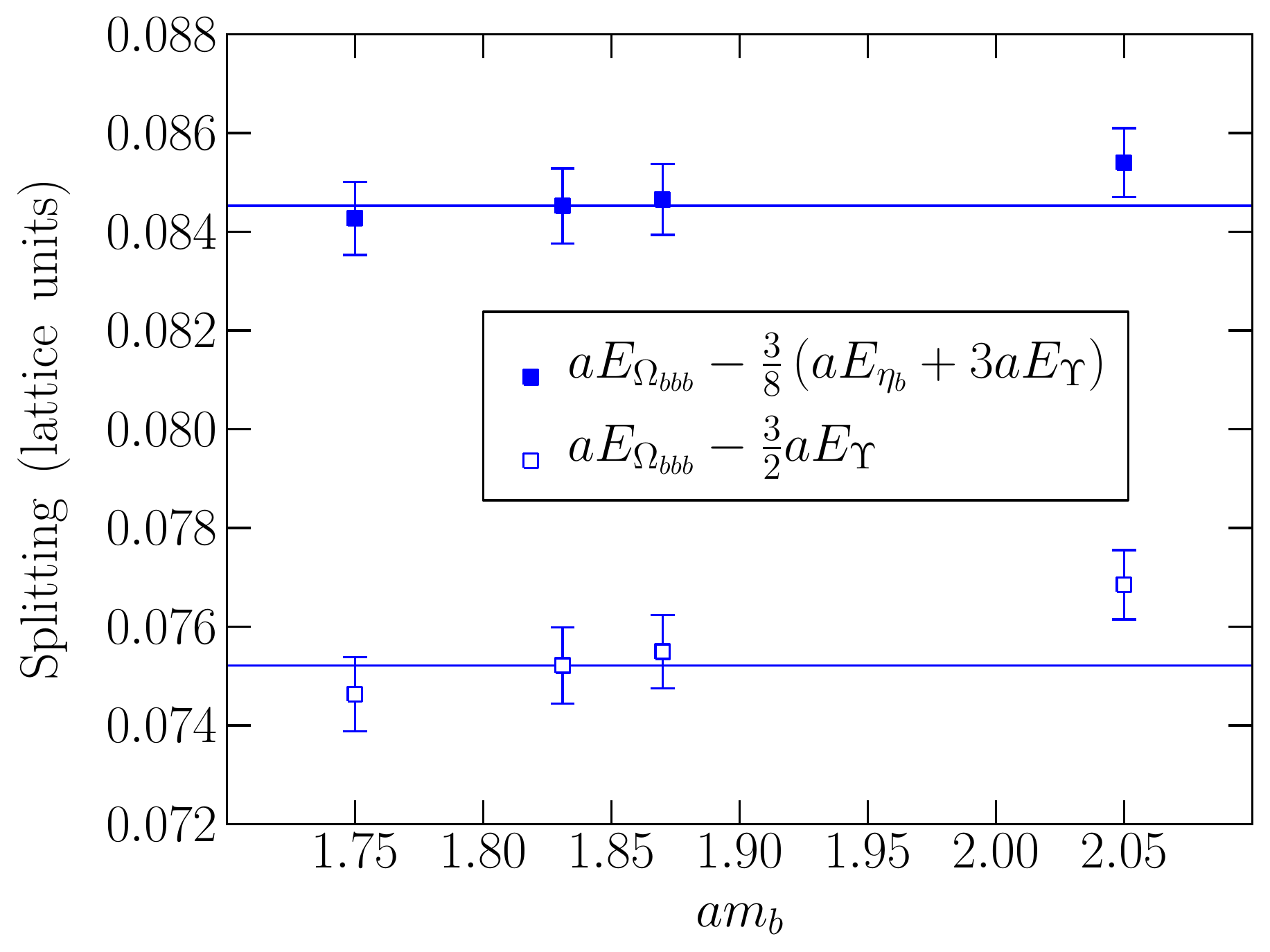}
\caption{\label{fig:Omega_bbb_vs_mb}Dependence of the quantities $aE_{\Omega_{bbb}}-\frac38\left( aE_{\eta_b}+3 aE_{\Upsilon}\right)$ and $aE_{\Omega_{bbb}}- \frac32aE_{\Upsilon}$
on the bare heavy-quark mass. Left panel: RBC/UKQCD $L=24$, $a m_l=0.005$; right panel: RBC/UKQCD $L=32$, $a m_l=0.004$. The lines indicate the values at the physical $b$-quark mass.}
\end{figure*}

\begin{figure*}[h!]
 \includegraphics[width=0.43\linewidth]{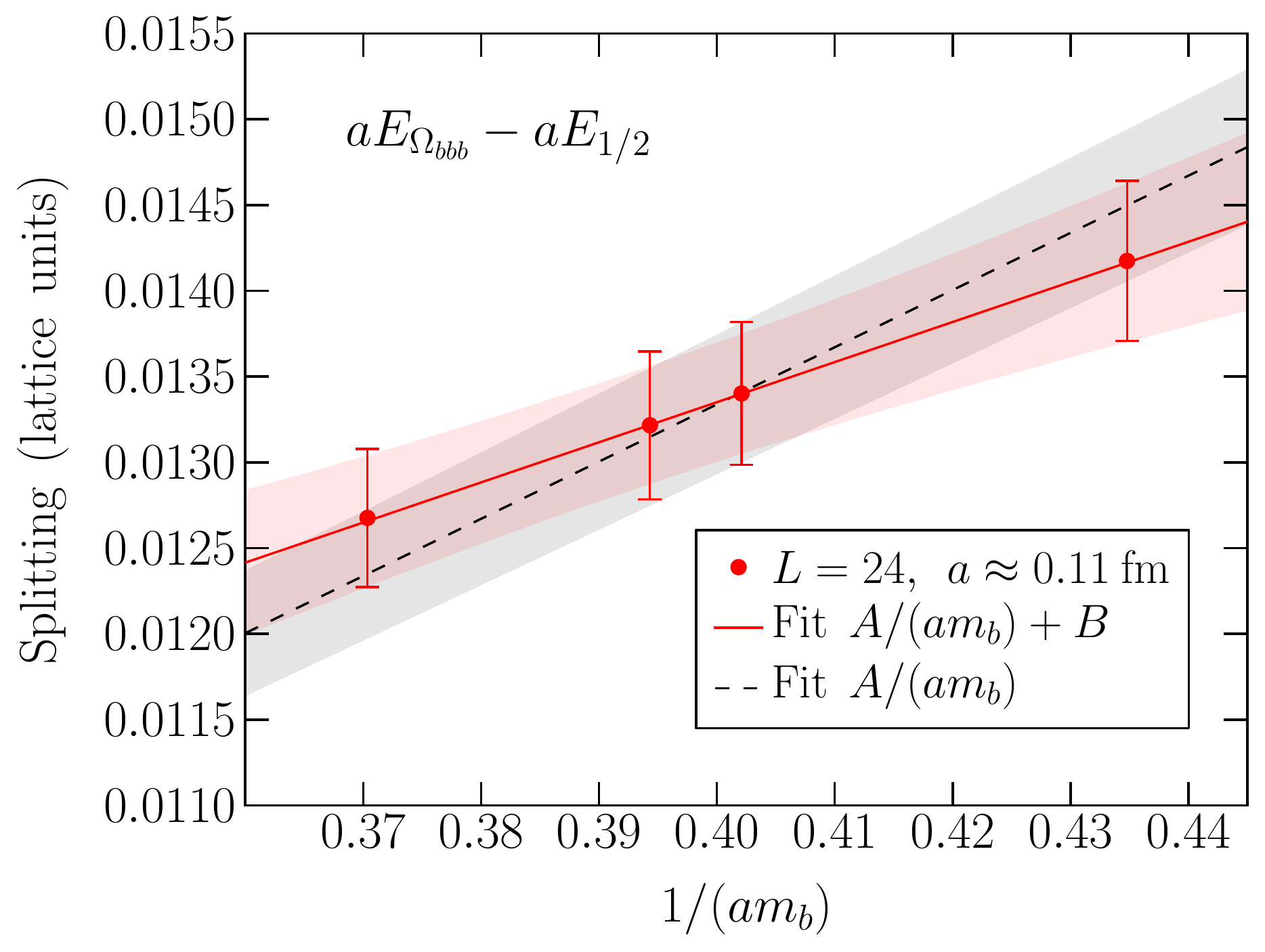} \hfill \includegraphics[width=0.43\linewidth]{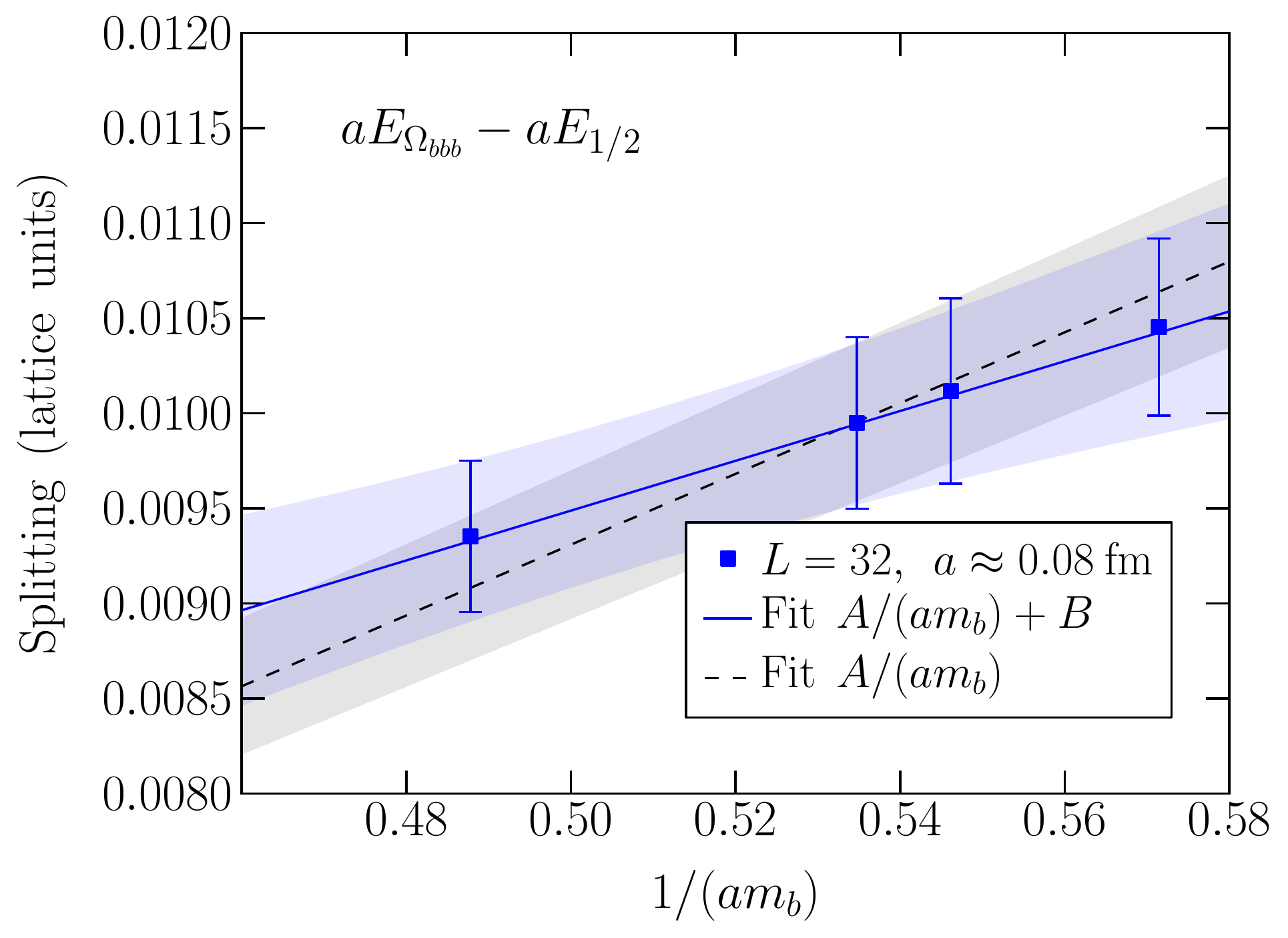}
\caption{\label{fig:Omega_bbb_spin_splitting_vs_mb}Dependence of the (unphysical) spin splitting $aE_{\Omega_{bbb}}-aE_{1/2}$ on the bare heavy-quark mass. Left panel: RBC/UKQCD $L=24$, $a m_l=0.005$;
right panel: RBC/UKQCD $L=32$, $a m_l=0.004$. Also shown are fits using the functions $A/(a m_b)$ and $A/(a m_b)+B$.}
\end{figure*}

The lattice results for the energy differences (\ref{eq:mdiff_spinav}), (\ref{eq:mdiff_1s}), and the unphysical spin splitting
$aE_{\Omega_{bbb}}-aE_{1/2}$ are given in Table \ref{tab:results_lattice_units}. On the RBC/UKQCD ensembles with $L=24$,
$a m_l=0.005$ and $L=32$, $a m_l=0.004$, results for multiple values of $a m_b$ are listed. The quark-mass-dependence of
(\ref{eq:mdiff_spinav}) and (\ref{eq:mdiff_1s}) is visualized in Fig.~\ref{fig:Omega_bbb_vs_mb}. As can be seen
there, the dependence on $a m_b$ is smaller for (\ref{eq:mdiff_spinav}). The quark-mass-dependence of the hyperfine splitting 
$aE_{\Omega_{bbb}}-aE_{1/2}$ is visualized in Fig.~\ref{fig:Omega_bbb_spin_splitting_vs_mb}. As can be seen
in the plot, the dependence is slightly weaker than $1/(a m_b)$, similarly to the $1S$ hyperfine splitting in
bottomonium \cite{Meinel:2010pv}. The data are described well by fits with the function $A/(a m_b)+B$.

On the $L=24$, $a m_l=0.005$ and $L=32$, $a m_l=0.004$ ensembles, results calculated with the leading-order (order-$v^2$) NRQCD action
are also given in Table \ref{tab:results_lattice_units}. This action does not contain spin-dependent terms, so that $aE_{\Omega_{bbb}}-aE_{1/2}=0$
and $a E_\Upsilon-a E_{\eta_b}=0$, and consequently the quantities (\ref{eq:mdiff_spinav}) and (\ref{eq:mdiff_1s}) become equal.
Remarkably, for the splitting $aE_{\Omega_{bbb}}-\frac38\left( aE_{\eta_b}+3 aE_{\Upsilon}\right)$ the results obtained
at the same value of $a m_b$ with the order-$v^4$ and with the order-$v^2$ actions differ by less than 1\%. It appears that
there is a large cancellation of the order-$v^4$ corrections in $aE_{\Omega_{bbb}}$ and $\frac38\left( aE_{\eta_b}+3 aE_{\Upsilon}\right)$,
which is another reason to use (\ref{eq:mdiff_spinav}) for the determination of $M_{\Omega_{bbb}}$.
On the other hand, for the splitting $aE_{\Omega_{bbb}}- \frac32aE_{\Upsilon}$, the values obtained with the
order-$v^4$ and with the order-$v^2$ actions differ by about 10\%, as might be expected for $v^2\approx 0.1$.

\begin{figure*}[h!]
 \includegraphics[width=0.43\linewidth]{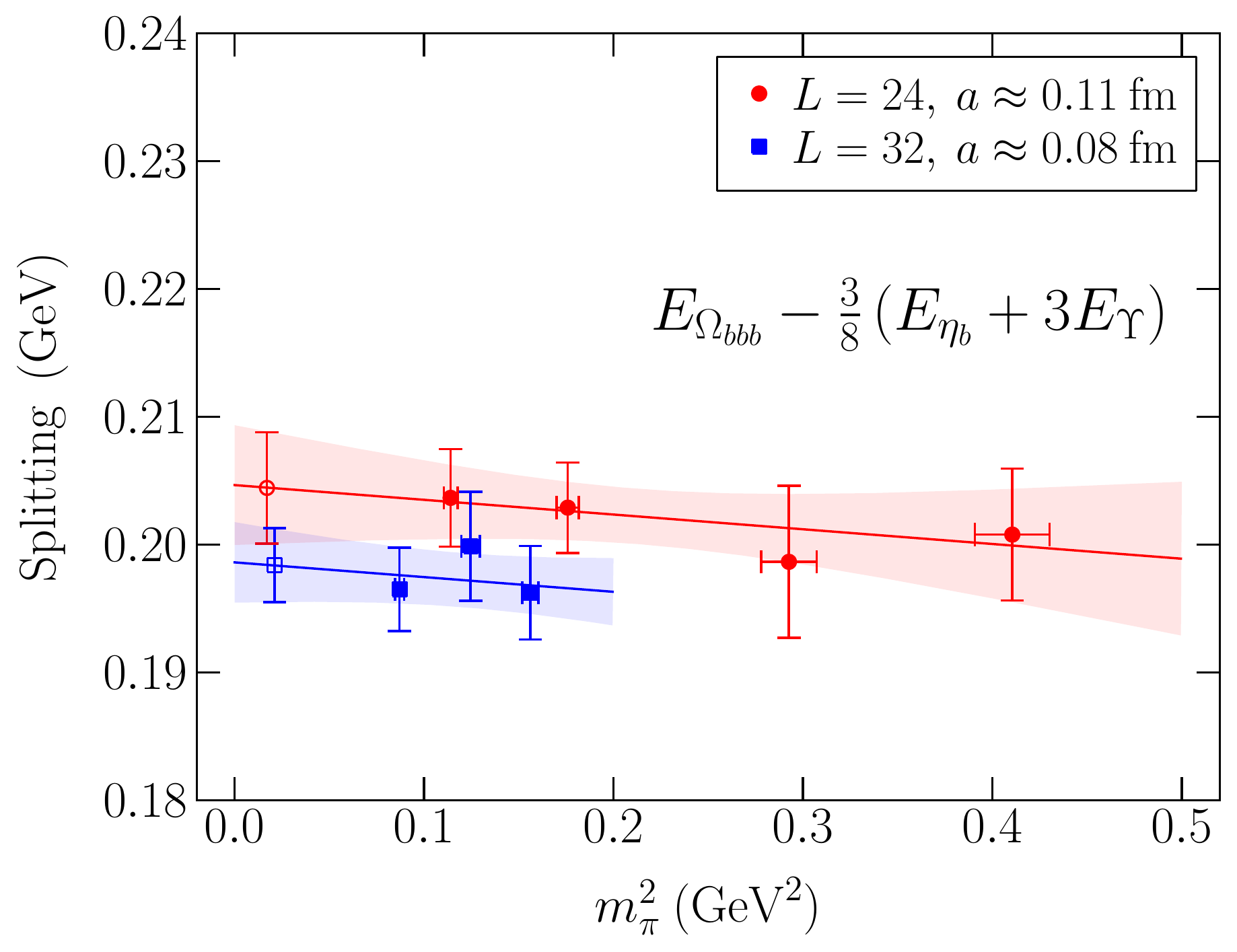} \hfill \includegraphics[width=0.43\linewidth]{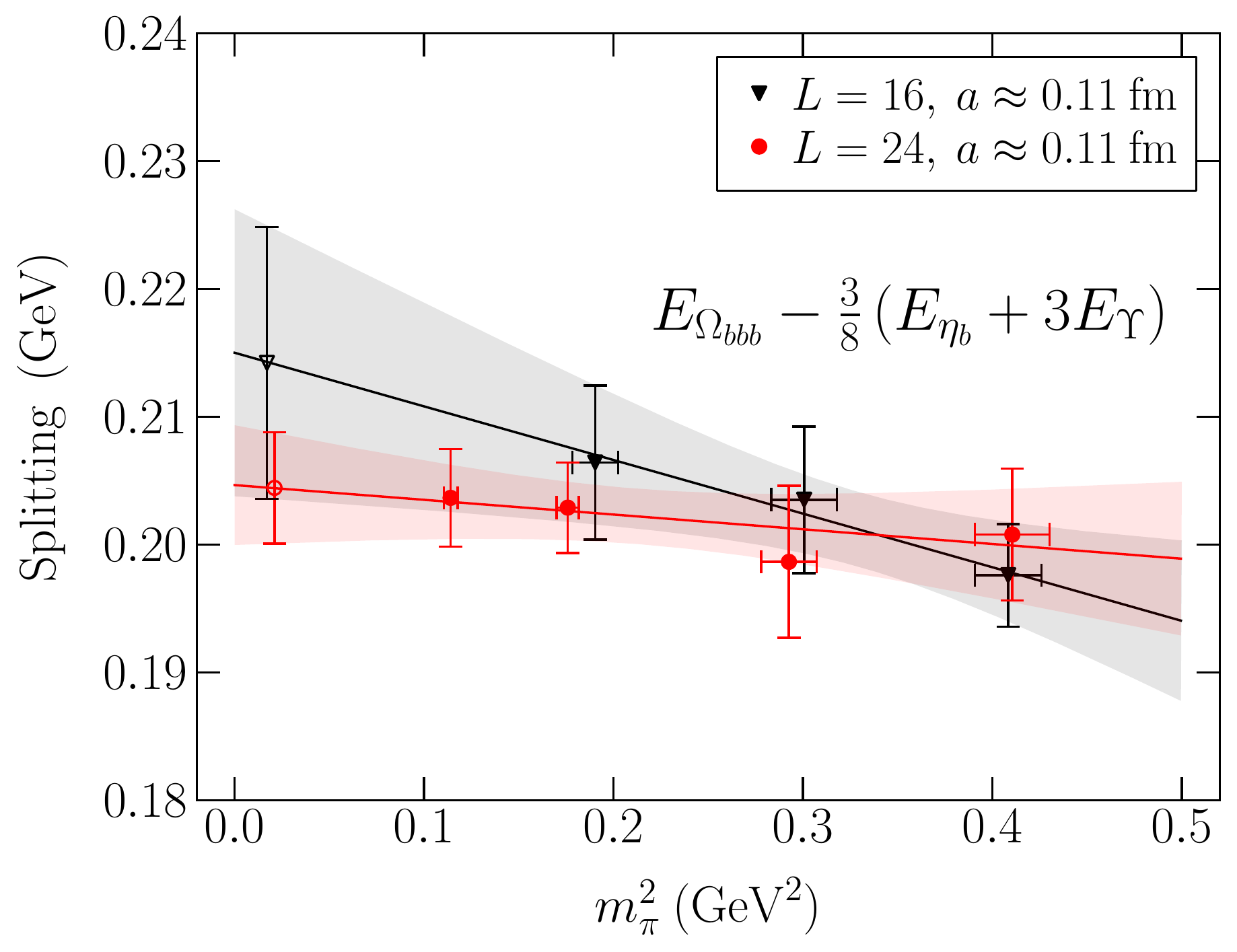}
\caption{\label{fig:Omega_bbb_1Sav_chiral_extrap}Chiral extrapolation of the quantity $E_{\Omega_{bbb}}-\frac38\left( E_{\eta_b}+3 E_{\Upsilon}\right)$
from the RBC/UKQCD gauge field ensembles. The extrapolated points are offset horizontally for legibility.}
\end{figure*}

\begin{figure*}[h!]
 \includegraphics[width=0.43\linewidth]{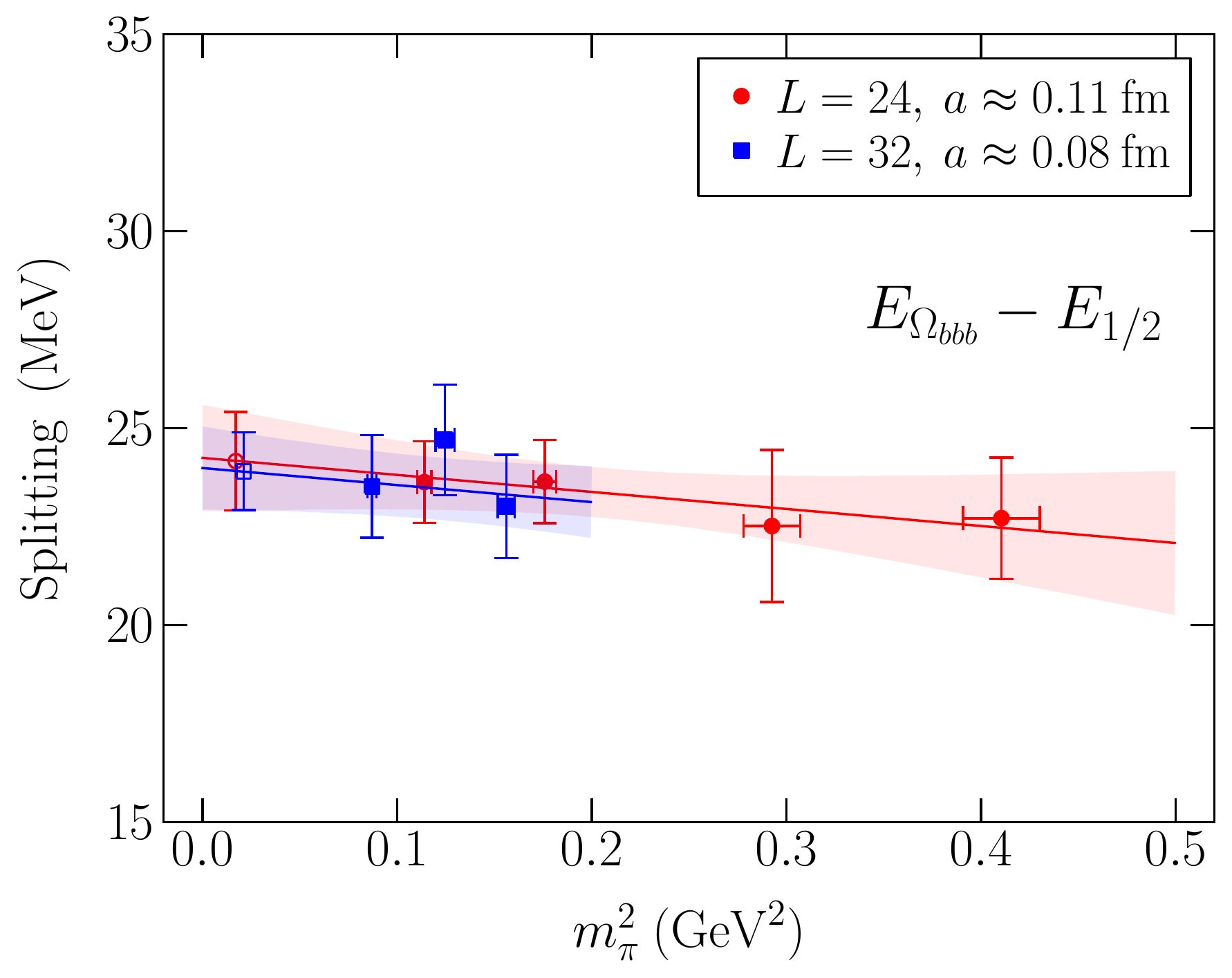} \hfill \includegraphics[width=0.43\linewidth]{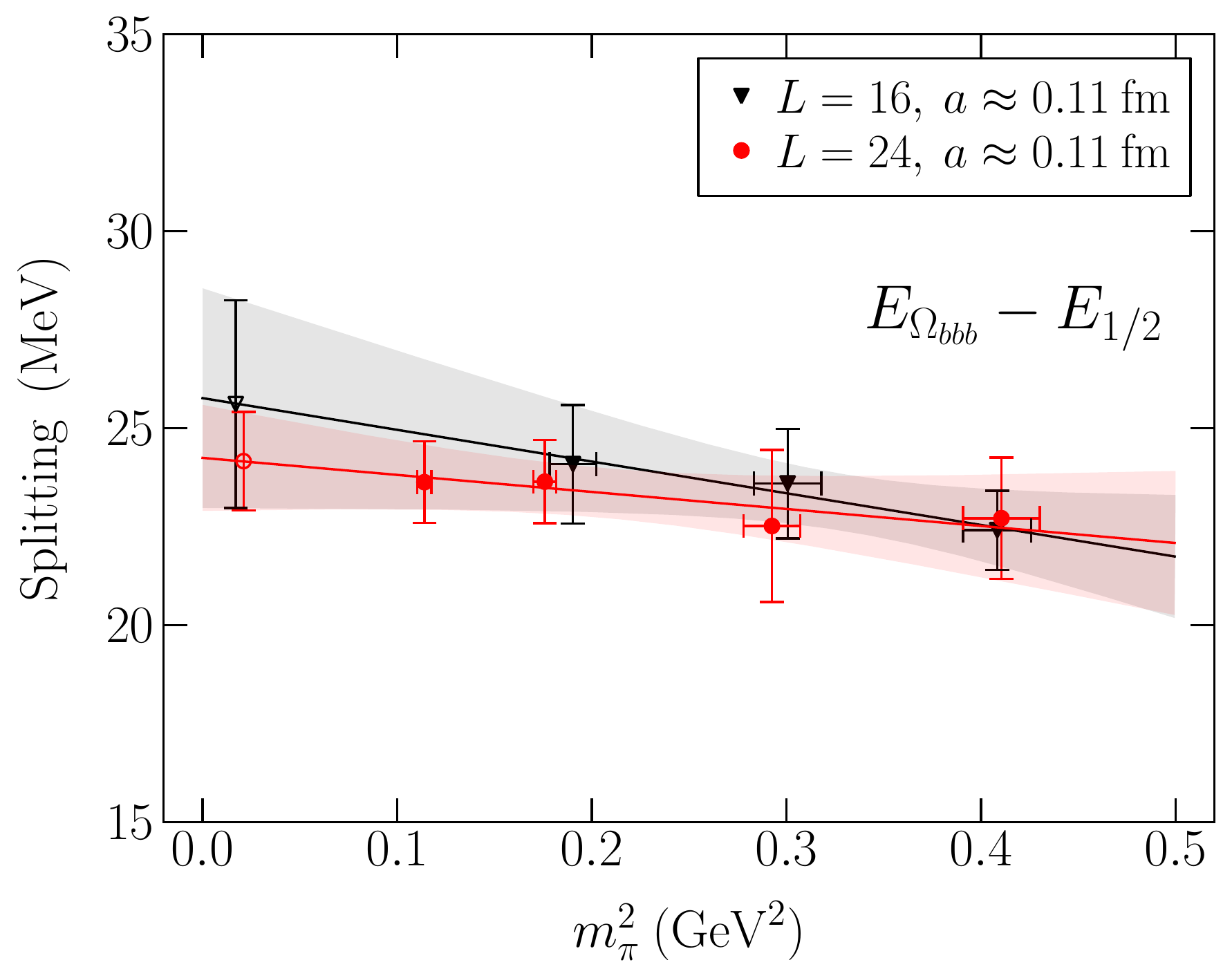}
\caption{\label{fig:Omega_bbb_spin_splitting_chiral_extrap}Chiral extrapolation of the (unphysical) spin splitting $E_{\Omega_{bbb}}-E_{1/2}$ from the RBC/UKQCD gauge field ensembles.
Extrapolated points are offset horizontally for legibility.}
\end{figure*}

The results for $aE_{\Omega_{bbb}}-\frac38\left( aE_{\eta_b}+3 aE_{\Upsilon}\right)$ and $aE_{\Omega_{bbb}}-aE_{1/2}$
from the order-$v^4$ action at the physical values of $a m_b$ were then converted to physical
units using the inverse lattice spacing values from Table \ref{tab:lattices}. The chiral extrapolations of the data
from the RBC/UKQCD ensembles are visualized in Figs.~\ref{fig:Omega_bbb_1Sav_chiral_extrap} and
\ref{fig:Omega_bbb_spin_splitting_chiral_extrap}. As can be seen there,
the dependence of the results on the pion mass is weak. This also
indicates that the higher-order terms depending on both $a$ and $m_\pi$ are small.
The data from the $L=16$ ensembles, which provide a test of finite-volume effects,
were extrapolated independently. No significant finite-volume effects are seen,
as expected for the small size of the $\Omega_{bbb}$ and bottomonium ground states
(Eqs.~(\ref{eq:size_omega}) and (\ref{eq:size_upsilon})).

\vspace{3ex}

\begin{table*}[h!]
\begin{ruledtabular}
\begin{tabular}{lcllcccc}
Collaboration & $L^3\times T$ & $a m_l$ & $a m_b$ & Action & $aE_{\Omega_{bbb}}\!\!-\frac38\left( aE_{\eta_b}\!\!+3 aE_{\Upsilon}\right)$ & $aE_{\Omega_{bbb}}\!\!- \frac32aE_{\Upsilon}$ & $aE_{\Omega_{bbb}}-aE_{1/2}$ \\
\hline
RBC/UKQCD & $16^3\times 32$  & $0.01$   & $2.469$ & $v^4$ & $0.11690(65)$   & $0.10527(64)$   &  $0.01364(30)$  \\
          & $16^3\times 32$  & $0.02$   & $2.604$ & $v^4$ & $0.12063(65)$   & $0.10934(64)$   &  $0.01399(31)$  \\
          & $16^3\times 32$  & $0.03$   & $2.689$ & $v^4$ & $0.11972(57)$   & $0.10856(56)$   &  $0.01357(28)$  \\
\\[-1ex]
RBC/UKQCD & $24^3\times 64$  & $0.005$  & $2.3$   & $v^4$ & $0.11483(95)$   & $0.10269(96)$   &  $0.01417(47)$  \\
          & $24^3\times 64$  & $0.005$  & $2.487$ & $v^4$ & $0.11551(97)$   & $0.10407(98)$   &  $0.01340(42)$  \\
          & $24^3\times 64$  & $0.005$  & $2.487$ & $v^2$ & $0.11445(78)$   & $0.11445(78)$   &  $0$            \\
          & $24^3\times 64$  & $0.005$  & $2.536$ & $v^4$ & $0.11568(99)$   & $0.10442(98)$   &  $0.01322(43)$  \\
          & $24^3\times 64$  & $0.005$  & $2.7$   & $v^4$ & $0.11631(95)$   & $0.10566(93)$   &  $0.01268(40)$  \\
          & $24^3\times 64$  & $0.01$   & $2.522$ & $v^4$ & $0.11731(78)$   & $0.10586(81)$   &  $0.01365(42)$  \\
          & $24^3\times 64$  & $0.02$   & $2.622$ & $v^4$ & $0.1183(17)\nb$ & $0.1072(17)\nb$ &  $0.01343(94)$  \\
          & $24^3\times 64$  & $0.03$   & $2.691$ & $v^4$ & $0.1217(14)\nb$ & $0.1106(14)\nb$ &  $0.01377(66)$  \\
\\[-1ex]
RBC/UKQCD & $32^3\times 64$  & $0.004$  & $1.75$  & $v^4$ & $0.08427(74)$   & $0.07463(75)$   &  $0.01045(47)$  \\
          & $32^3\times 64$  & $0.004$  & $1.831$ & $v^4$ & $0.08453(76)$   & $0.07521(77)$   &  $0.01012(49)$  \\
          & $32^3\times 64$  & $0.004$  & $1.831$ & $v^2$ & $0.08473(48)$   & $0.08473(48)$   &  $0$            \\
          & $32^3\times 64$  & $0.004$  & $1.87$  & $v^4$ & $0.08466(72)$   & $0.07549(74)$   &  $0.00995(45)$  \\
          & $32^3\times 64$  & $0.004$  & $2.05$  & $v^4$ & $0.08540(70)$   & $0.07685(70)$   &  $0.00935(40)$  \\
          & $32^3\times 64$  & $0.006$  & $1.829$ & $v^4$ & $0.08581(68)$   & $0.07647(68)$   & $0.01061(44)$  \\
          & $32^3\times 64$  & $0.008$  & $1.864$ & $v^4$ & $0.08584(98)$   & $0.07516(98)$   & $0.01007(50)$  \\
\\[-1ex]
MILC      & $24^3\times 64$  & $0.005$  & $2.64$  & $v^4$ & $0.12447(64)$   & $0.11216(64)$   & $0.01438(30)$  \\
\\[-1ex]
MILC      & $28^3\times 96$  & $0.0062$ & $1.86$  & $v^4$ & $0.08757(49)$   & $0.07786(50)$   & $0.01022(24)$  \\
\end{tabular}
\end{ruledtabular}
\caption{\label{tab:results_lattice_units}Results in lattice units. The errors are statistical/fitting only.}
\end{table*}

\vspace{2ex}

The results of the chiral extrapolations of $E_{\Omega_{bbb}}-\frac38\left( E_{\eta_b}+3 E_{\Upsilon}\right)$ and $E_{\Omega_{bbb}}-E_{1/2}$ to $m_\pi=138$ MeV
are given in Table \ref{tab:extrapolated_results}. Also shown in the table are the results from the MILC ensembles, and
the results from the $L=24$ and $L=32$ RBC/UKQCD ensembles after interpolation/extrapolation to $m_\pi=460$ MeV
and $m_\pi=416$ MeV, respectively, so as to match the root-mean-square pion masses of the MILC ensembles.
As mentioned above, on the coarse MILC ensemble the value of $a m_b$ used here is 0.9\% below the physical value obtained in
\cite{Meinel:2010pv}. However, this affects the hyperfine splitting in physical units by only 0.2 standard deviations, which is negligible.

For the hyperfine splitting, no significant dependence on the lattice spacing is seen, and the results from the RBC/UKQCD and MILC ensembles at the matching pion
masses are in agreement, indicating that gluon discretization errors for this quantity are small.
However, for this splitting systematic errors of order $\alpha_s\approx20\hdots30$\% and $v^2\approx 10$\% due to missing radiative and relativistic corrections in the NRQCD action
are expected. The final result for the (fictitious) hyperfine splitting is then
\begin{equation}
E_{\Omega_{bbb}}-E_{1/2}=24\pm 1_{\rm\:stat}\pm 6_{\rm\:syst}\:\:{\rm MeV}.
\end{equation}

The quantity $E_{\Omega_{bbb}}-\frac38\left( E_{\eta_b}+3 E_{\Upsilon}\right)$ is seen to change by about 6 MeV (1.1$\sigma$) when going from the coarse to the fine lattice spacing
on the RBC/UKQCD ensembles. At the coarse lattice spacing and at the matching pion masses, the results from the RBC/UKQCD and MILC ensembles
differ by about 3 MeV (1.0$\sigma$); at the fine lattice spacing this difference is 4 MeV (1.2$\sigma$). Given that the ratio of $a^2$ between the fine and coarse lattices is about 2,
it seems reasonable to assume a maximum total discretization error of 10 MeV (5\%) for the result from the fine ($L=32$) RBC/UKQCD ensemble. The systematic errors from the missing radiative
and relativistic corrections in the NRQCD action can be estimated to be of order $\alpha_s v^2\approx2\hdots3$\% and $v^4\approx1$\% respectively. Thus, the final result
for $E_{\Omega_{bbb}}-\frac38\left( E_{\eta_b}+3 E_{\Upsilon}\right)$ (without the electrostatic correction to be discussed in Sec.~\ref{sec:mass}) is
\begin{equation}
 E_{\Omega_{bbb}}-\frac38\left( E_{\eta_b}+3 E_{\Upsilon}\right)=0.198\pm0.003_{\rm \:stat} \pm 0.011_{\rm \:syst}\:\:{\rm GeV}. \label{eq:mass_splitting_final}
\end{equation}

\begin{table}[h!]
\begin{tabular}{lcccccccc}
\hline\hline
Collaboration & \hspace{4ex} & $L^3\times T$ & \hspace{4ex} & $m_\pi$ (GeV) & \hspace{4ex} & $E_{\Omega_{bbb}}-\frac38\left( E_{\eta_b}+3 E_{\Upsilon}\right)$ (GeV) & \hspace{4ex} & $E_{\Omega_{bbb}}-E_{1/2}$ (MeV) \\
\hline
RBC/UKQCD && $16^3\times 32$ && $0.138$ && $0.214(11)\nb$ && $25.6(2.6)\nb$ \\
RBC/UKQCD && $24^3\times 64$ && $0.138$ && $0.2044(44)$   && $24.2(1.2)\nb$ \\
RBC/UKQCD && $32^3\times 64$ && $0.138$ && $0.1984(29)$   && $23.91(99)\phantom{.}$ \\
\\[-1ex]
MILC      && $24^3\times 64$ && $0.460$ && $0.2052(22)$   && $23.69(65)\phantom{.}$ \\
RBC/UKQCD && $24^3\times 64$ && $0.460$ && $0.2022(22)$   && $23.33(63)\phantom{.}$ \\
\\[-1ex]
MILC      && $28^3\times 96$ && $0.416$ && $0.2008(24)$   && $23.41(72)\phantom{.}$ \\
RBC/UKQCD && $32^3\times 64$ && $0.416$ && $0.1966(24)$   && $23.24(84)\phantom{.}$ \\
\hline\hline
\end{tabular}
\caption{\label{tab:extrapolated_results}Values of $E_{\Omega_{bbb}}-\frac38\left( E_{\eta_b}+3 E_{\Upsilon}\right)$ and $E_{\Omega_{bbb}}-E_{1/2}$ in physical units.
The errors are statistical/fitting only. The first three rows of the table give the results from the RBC/UKQCD ensembles, extrapolated to the physical pion mass.
In the remaining rows, the results from the MILC ensembles are compared to the results from the RBC/UKQCD ensembles; there, the RBC/UKQCD data were interpolated/extrapolated
to match the root-mean-square pion masses of the MILC ensembles.}
\end{table}

\section{\label{sec:mass}The mass of the $\Omega_{bbb}$}

The result (\ref{eq:mass_splitting_final}) is the pure QCD value and needs to be corrected to include the effects of electromagnetism.
Since the quarks in both the $\Omega_{bbb}$ and bottomonium are heavy and are moving slowly, the dominant electromagnetic correction is due to the
electrostatic Coulomb interaction. The Coulomb interaction is repulsive in the $\Omega_{bbb}$
and it therefore increases $E_{\Omega_{bbb}}$. On the other hand, the Coulomb interaction is attractive in bottomonium and it therefore
decreases $E_{\eta_b}$ and $E_{\Upsilon}$. Using $\langle \Upsilon | r^{-1} | \Upsilon \rangle=\langle \eta_b | r^{-1} | \eta_b \rangle$,
the electrostatic correction to (\ref{eq:mass_splitting_final}) becomes
\begin{equation}
 E_{\rm Coulomb} = 3 \frac{(e/3)^2}{4\pi\epsilon_0} \langle \Omega_{bbb} | \frac1r | \Omega_{bbb} \rangle
+ \frac 32 \frac{(e/3)^2}{4\pi\epsilon_0} \langle \Upsilon | \frac1r | \Upsilon \rangle. \label{eq:electrostatic_correction}
\end{equation}
Here, $r$ is defined as the distance between two $b$ quarks (for the $\Omega_{bbb}$) or the distance between the $b$ and $\overline{b}$ (for the $\Upsilon$).
The expectation value $\langle \Upsilon | r^{-1} | \Upsilon \rangle$ is calculated numerically using the $1S$ wave function obtained with the
QQ-onia package \cite{DomenechGarret:2008sr} for the Cornell potential with parameters as in \cite{DomenechGarret:2008sr}. This gives
\begin{equation}
 \langle \Upsilon | \frac1r | \Upsilon \rangle = 8.1\:\:{\rm fm}^{-1}.
\end{equation}
In Ref.~\cite{SilvestreBrac:1996bg}, the mean-square distance of a heavy quark from the center of mass in the $\Omega_{bbb}$ has been calculated using a potential model,
with the result $\langle \Omega_{bbb} | r^2_{\rm CM} | \Omega_{bbb} \rangle = 0.021 \:\:{\rm fm}^2$. The geometry of the system implies $r=\sqrt{3} \: r_{\rm CM}$,
and therefore
\begin{equation}
\sqrt{\langle \Omega_{bbb} | r^2 | \Omega_{bbb} \rangle}=0.25 \:\:{\rm fm}. \label{eq:size_omega}
\end{equation}
For comparison, in bottomonium the wave function from QQ-onia gives
\begin{equation}
\sqrt{\langle \Upsilon | r^2 | \Upsilon \rangle}=0.20 \:\:{\rm fm}. \label{eq:size_upsilon}
\end{equation}
Therefore, in the following the estimate $\langle \Omega_{bbb} | r^{-1} | \Omega_{bbb} \rangle = (0.8\pm 0.4)\langle \Upsilon | r^{-1} | \Upsilon \rangle = 6.5\pm3.2\:\:{\rm fm}^{-1}$ is used.
This gives
\begin{equation}
 E_{\rm Coulomb} = 5.1 \pm 2.5 \:\:{\rm MeV}. \label{eq:electrostatic_correction_numerical}
\end{equation}
The full mass of the $\Omega_{bbb}$ is then calculated as
\begin{equation}
M_{\Omega_{bbb}} = \left[ E_{\Omega_{bbb}}-\frac38\left( E_{\eta_b}+3 E_{\Upsilon}\right) \right]_{\rm LQCD} + E_{\rm Coulomb} + \frac 3 2  \Big[ M_\Upsilon \Big]_{\rm PDG} - \frac 38 \Big[ E_{\Upsilon}-E_{\eta_b} \Big]. \label{eq:omega_bbb_full_mass}
\end{equation}
Here, the first term is given by (\ref{eq:mass_splitting_final}), and the last term (the bottomonium hyperfine splitting)
is taken from the lattice calculation \cite{Meinel:2010pv}:
\begin{eqnarray}
\nonumber E_{\Upsilon}-E_{\eta_b} &=&\left[ \frac{E_{\Upsilon}-E_{\eta_b}}{1P\:\:{\rm tensor}} \right]_{\rm LQCD}\cdot \Big[ 1P\:\:{\rm tensor} \Big]_{\rm PDG} \\
\nonumber && \\
&=&60.3\pm5.5_{\rm\:stat } \pm5.0_{\rm\: syst } \pm2.1_{\rm\: exp }\:\:{\rm MeV}. \label{eq:hyperfine}
\end{eqnarray}
In (\ref{eq:hyperfine}), the ratio of the $1S$ hyperfine splitting and the $1P$ tensor splitting is used \cite{Meinel:2010pv}. The experimental
uncertainty in the $1P$ tensor splitting \cite{Amsler:2008zzb} leads to the last error in (\ref{eq:hyperfine}). In Eq. (\ref{eq:omega_bbb_full_mass}), the mass of the $\Upsilon$
is taken from the Particle Data Group to be $M_\Upsilon=9.4603\pm0.0003$ GeV \cite{Amsler:2008zzb}. The final result for the mass
of the $\Omega_{bbb}$ is then
\begin{equation}
M_{\Omega_{bbb}} = 14.371 \pm 0.004_{\rm \:stat} \pm 0.011_{\rm \:syst} \pm 0.001_{\rm \:exp}\:\:{\rm GeV}. \label{eq:m_omega_final}
\end{equation}

\section{Conclusions}

The value for $M_{\Omega_{bbb}}$ obtained here, Eq.~(\ref{eq:m_omega_final}), is compared to results from various continuum calculations in Table \ref{tab:Omega_bbb_comparison}.
The results from the literature range from $13.28 \pm 0.10$ GeV, calculated using sum rules in \cite{Zhang:2009re}, to $14.834$ GeV calculated using a quark model
in \cite{Roberts:2007ni}. The nonperturbative dynamical lattice QCD calculation performed here, with a total
uncertainty of only 12 MeV, is a valuable test of the continuum models. If an experimental result for $M_{\Omega_{bbb}}$ ever becomes available,
it will in turn become a stringent test of the lattice calculation.

The value (\ref{eq:m_omega_final}) obtained here satisfies the
baryon-meson mass inequality $M_{\Omega_{bbb}}\geq \frac32 M_{\Upsilon}$ derived in
\cite{Ader:1981db,Nussinov:1983vh,Richard:1984wy}, like most of the results given in Table \ref{tab:Omega_bbb_comparison},
with the exception of those from \cite{Tsuge:1985ei} and \cite{Zhang:2009re}.

\begin{table}[h!]
\begin{center}
\begin{tabular}{lll}
\hline\hline
Reference                 & \hspace{2ex} & $M_{\Omega_{bbb}}$ (GeV) \\
\hline
Ponce \cite{Ponce:1978gk}                            & & $14.248$ \\
Hasenfratz {\it et al.} \cite{Hasenfratz:1980ka}     & & $14.30$ \\
Bjorken \cite{Bjorken:1985ei}                        & & $14.76 \pm 0.18$ \\
Tsuge {\it et al.} \cite{Tsuge:1985ei}               & & $13.823$ \\
Silvestre-Brac \cite{SilvestreBrac:1996bg}           & & $14.348 \hdots 14.398$ \\
Jia \cite{Jia:2006gw}                                & & $14.37 \pm 0.08$ \\
Martynenko \cite{Martynenko:2007je}                  & & $14.569$ \\
Roberts and Pervin \cite{Roberts:2007ni}             & & $14.834$ \\
Bernotas and Simonis \cite{Bernotas:2008bu}          & & $14.276$ \\
Zhang and Huang \cite{Zhang:2009re}                  & & $13.28 \pm 0.10$ \\
\\[-1ex]
This work                                            & & $14.371 \pm 0.004_{\rm \:stat} \pm 0.011_{\rm \:syst} \pm 0.001_{\rm \:exp}$ \\
\hline\hline
\end{tabular}
\caption{\label{tab:Omega_bbb_comparison}Comparison of results for the $\Omega_{bbb}$ mass.}
\end{center}
\end{table}

While the lattice calculation of $M_{\Omega_{bbb}}$ can \emph{rule out} many continuum models, the sole agreement for this value
from a model-dependent calculation with several parameters is of course not sufficient to show that the assumptions of that model are correct.
To establish the usefulness of a model, multiple predictions using only a small number of parameters need to be tested.
It is planned to perform lattice calculations also for excited states of the $\Omega_{bbb}$, which will give
more insight into the three-quark forces of QCD.
Excited states have been studied using a quark model in \cite{SilvestreBrac:1996bg}. Another direction
is the lattice calculation of masses of triply-heavy baryons containing charm quarks: $ccc$, $ccb$ and $cbb$.
Lattice NRQCD is not well suited for charm quarks due to their lower mass; instead, for example the Fermilab method
\cite{ElKhadra:1996mp} can be used for them.

\begin{acknowledgments}
I thank William Detmold for useful discussions. I thank the RBC/UKQCD and MILC collaborations for making their gauge field ensembles available.
This work was supported by the U.S.~Department of Energy under grant number {D}{E}-{S}{C00}01{784}.
The computations were performed using resources at Fermilab, NERSC, and Teragrid resources at NCSA (grant number TG-PHY080014N).
\end{acknowledgments}

\end{document}